\documentstyle[11pt]{article}

\newcommand{\ba}{\begin{array}}
\newcommand{\ea}{\end{array}}
\newcommand{\be}{\begin{equation}}
\newcommand{\ee}{\end{equation}}
\newcommand{\bea}{\begin{eqnarray}}
\newcommand{\eea}{\end{eqnarray}}
\newcommand{\beas}{\begin{eqnarray*}}
\newcommand{\eeas}{\end{eqnarray*}}

\def\wt{\widetilde}

\def\sect{\section}
\def\EQ{\begin{equation}}
\def\EN{\end{equation}}
\def\bea{\begin{eqnarray}}
\def\ena{\end{eqnarray}}
\newcommand{\vs}[1]{\vspace{#1 mm}}

\newcommand{\mathbb}[1]{{\bf{#1}}}

\begin{document}
\begin{titlepage}
\begin{center}

\hfill  {\sc Pre-Print\ CTP$\sharp$2887} \\
\hfill  {\sc July 1999} \\
\hfill  {\tt hep-th/yymmddd} \\
        [.35in]

{\large\bf
MATRIX THEORY STAR PRODUCTS FROM THE BORN--INFELD ACTION }\\
\medskip

{\bf Lorenzo Cornalba}\footnote{E-mail: {\tt cornalba@princeton.edu}  -- 
After Oct.~1st: {\it I.H.E.S., Le Bois-Marie, Bures-Sur-Yvette, 
91440 France}} \\
{\it Department of Physics}\\
{\it Joseph Henry Laboratories, Princeton University}\\
{\it Princeton, NJ 08544, U.S.A.}\\

\medskip
{\sl and}\\
\medskip

{\bf Ricardo Schiappa}\footnote{E-mail: {\tt ricardos@ctp.mit.edu}  --  
After Sep.~1st: {\it Dept. Physics, Harvard University, Cambridge, MA 02138}} \\
{\it Center for Theoretical Physics and Department of Physics}\\
{\it Massachusetts Institute of Technology, 77 Massachusetts Ave.}\\
{\it Cambridge, MA 02139, U.S.A.}\\

\end{center}

\vs{4}
\centerline{{\bf{Abstract}}}
\vs{4}

We conjecture that the Sen--Seiberg limit of the Type IIA $D2$--brane action in a flat 
spacetime background can be resummed, at all orders in $\alpha'$, to define an 
associative star product on the membrane. This star product can be independently constrained
from the equivalent Matrix theory description of the corresponding 
$M2$--brane, by carefully analyzing the known BPS conditions. Higher derivative corrections 
to the Born--Infeld action on the IIA side are reinterpreted, after the Sen--Seiberg limit, as 
higher derivative corrections to a field theory on the membrane, which itself can be resummed 
to yield the known Matrix theory quantum mechanics action. Conversely, given the 
star product on the membrane as a formal power series in $\alpha'$, one can constrain the higher 
derivative corrections to the Born--Infeld action, in the Sen--Seiberg limit. This claim is 
explicitly verified to first order. Finally, we also comment on the possible application of this
method to the derivation of the Matrix theory action for membranes in a curved background.
\end{titlepage}

\newpage
\renewcommand{\thefootnote}{\arabic{footnote}}
\setcounter{footnote}{0}

\sect{Introduction and Discussion}

It is by now clear that string theory as it emerged in the eighties is but a fraction of 
the full story. The five known superstring theories as well as the low--energy 11--dimensional 
supergravity are now known to be related through a web of dualities \cite{Witten1}, and it is believed 
that all these theories are simply different limits 
of an underlying 11--dimensional quantum theory known as $M$--theory, whose fundamental 
degrees of freedom are as yet unknown. Throughout these string theories we have $D$--branes of 
various dimensions \cite{Polchinsky}, 
playing in certain regimes a role as fundamental as that of the basic string. 
These $D$--branes have a precise description within string theory, their 
low--energy dynamics being dictated by the Born--Infeld action \cite{Leigh} 
which can be obtained using boundary 
conformal field theory techniques. $M$--theory itself contains $M2$--branes and $M5$--branes, 
which themselves can be related to the branes in string theories \cite{BFSS}.

The problem of understanding the structure of string theory, is directly related to a full
understanding of what is the fundamental nature of spacetime. At large distances the structure of the 
spacetime manifold is well described by Riemannian geometry. On the other hand, at small 
distances -- {\it i.e.}, much shorter than the string scale $\ell_s$ -- the Riemannian 
description of spacetime breaks down. In fact, there is as yet no known intrinsic and covariant 
description of the underlying geometry and fundamental degrees of freedom. Despite this lack 
of understanding of the small distance structure, one can use $D$--branes in order to 
probe physics at distances smaller than the string length \cite{DKPS}.

In this work, we shall focus our attention on the description of the $M$--theory $M2$--brane. 
Let us first recall that $M$--theory compactified on a circle is described 
by Type IIA string theory at finite string coupling \cite{Witten1}.
Moreover, the (unwrapped) $M2$--brane 
corresponds, in the Type IIA description, to the Dirichlet membrane. It is by now a well known 
conjecture that $M$--theory compactified on a light--like circle admits a non--perturbative 
description in terms of the degrees of freedom of a collection of $D0$--branes
\cite{BFSS,Susskind}. Moreover, 
these $D0$--branes are exactly described by the $(0+1)$--dimensional reduction of 10--dimensional 
$U(N)$ super Yang--Mills theory. A precise understanding of this Matrix theory conjecture 
is obtained by starting with $M$--theory, now compactified on a space--like circle, and 
considering the sector with $N$ units of momenta in the compact direction. From a IIA point of 
view this corresponds to the sector of the theory with $N$ $D0$--branes. One then 
uses the Poincar\'e invariance of the underlying $M$--theory in order to relate the 
space--like compactified theory to the theory compactified on the light--cone. This is achieved 
by both sending the space--like compactification radius to zero and, at the same time, by rescaling 
the $11$--dimensional Planck length. Such a prescription is known as the Sen--Seiberg limit
\cite{Sen,Seiberg}.

Membranes exist in both Type IIA string theory and in $M$--theory. In particular, one should be 
able to describe $M2$--branes as specific states within Matrix theory. On the other hand, 
$D2$--branes can be described within Type IIA string theory in terms of the geometry of a 
world--volume manifold embedded in spacetime, whose dynamics are governed at low energies 
by the Born--Infeld action and in general, at weak coupling, by the full $\alpha'$ expansion of the 
boundary state conformal field theory \cite{Leigh}.
Naively one would assume that these two descriptions should 
be essentially the same -- {\it i.e.}, they should coincide in the Sen--Seiberg limit. On the 
other hand, a closer look shows that these descriptions are actually valid in seemingly 
different regimes. In fact, Matrix theory probes the dynamics of the $D0$--branes at 
the Planckian length which, in the Sen--Seiberg limit, is much smaller than the 
string length. The $\alpha'$ expansion is, on the other hand,
a large distance expansion therefore valid 
in the opposite regime. Still, we shall conjecture in this work that one can learn 
about one description from the other. In particular, the exact knowledge of the Matrix theory 
action can be used to conjecture the Sen--Seiberg limit of the action describing the 
$D2$--brane to all orders in $\alpha'$ -- {\it i.e.}, Born--Infeld plus higher derivative 
corrections. Conversely, one could proceed the other way around and use extensions of 
known results \cite{BBG} in order to start tackling the problem of describing Matrix theory in curved 
backgrounds. Indeed one could use conformal field theory calculations in order to determine the 
$\alpha'$ corrections to the Born--Infeld action in a curved background. Then, via the prescription 
described in this paper, one could take the Sen--Seiberg limit and obtain the first corrections 
to Matrix theory in this curved background  \cite{TvR1,TvR2,KT,Douglas1,DKO}.

Let us now be a bit more specific. In order to carry out the above program, we shall start by 
considering static BPS membranes in Matrix theory. At the classical level, in flat 
background spacetime, these membranes are described by holomorphic curves. Using the work of 
Cornalba and Taylor \cite{CT}, one can represent any holomorphically embedded curve via matrices --
{\it i.e.}, we can construct its matrix representation. This construction is accomplished by 
starting with a given holomorphic embedding and then associating functions on the membrane 
to matrices, so that matrix multiplication is represented by an associative star product
on the space of functions. This star product is clearly non--commutative and, moreover, 
it is given by a formal power expansion in $\alpha'=\ell_P^3/R$, which at each order is 
given by a local derivative bilinear of the functions living on the membrane. The product starts 
as simple function multiplication, and is constrained at higher orders by the 
following three conditions. First, the star commutator of functions starts at order $\alpha'$ 
with the Poisson bracket, where the symplectic form is determined by the area element of 
the embedded curve. Secondly, and as previously stated, the product should be associative. 
Finally, the coordinate functions which represent at the Matrix level the static curve, 
should satisfy the BPS condition which follows from the known Matrix theory action. As described 
in \cite{CT} these three conditions can be iteratively solved (algebraically) in powers of 
$\alpha'$, if one considers the specific case of holomorphic quantization for the BPS membranes.

Having described the static BPS branes, we now proceed by considering fluctuations about 
these solutions. Fluctuations can again be represented dually either as matrices or as 
functions on the brane. In particular, if we consider the fluctuating coordinates as functions 
and we re--write the Matrix theory action in terms of the star product, one immediately 
gets an action in terms of higher derivatives of these coordinates functions 
and in powers of $\alpha'$. 
It is then very natural to conjecture that this expansion is nothing but the Sen--Seiberg 
limit of the full $\alpha'$ expansion describing the $D2$--brane.

What we do in this paper is, first of all, to describe in detail the correct Sen--Seiberg 
limit of the $D2$--brane action. We then apply this procedure to the Born--Infeld action 
and show that, in the limit, it reproduces exactly the light--cone gauge fixed Nambu--Goto 
$M2$--brane action in 11--dimensions. This is, as expected, the first term in the previously 
described expansion of the Matrix theory action as given in terms of the star product. 
Finally, we describe how one should proceed in order to test the conjecture at higher orders 
in $\alpha'$, and moreover how one can infer higher derivative terms in the $D2$--brane 
action from a given star product.

The picture that emerges from this conjecture is the following. Each derivative correction to 
the Born--Infeld action translates into a correction to the effective field theory on the 
$M2$--brane. The sum of all these derivative corrections on the $D2$--brane action translate 
to an infinite sum of corrections to the field theory on the $M2$--brane, in such a way that, 
in the Sen--Seiberg limit, they can be re--organized into a star product. This reduces the 
non--renormalizable field theory on the $M2$--brane to a matrix quantum mechanics. Moreover, 
this star product can be independently constrained by analyzing the BPS condition 
arising from Matrix theory.

\medskip

While this paper was being typed, we learned about other work along similar lines \cite{strings99}.

\sect{The Sen--Seiberg Limit of the Born--Infeld Action}

Our ultimate goal is to show how one can relate the action for an $M$--theory membrane 
to the action describing a $D2$--brane in Type IIA string theory. To this end, we have to
analyze the Sen--Seiberg limit and precisely understand how it relates 
states in light--cone compactified $M$--theory to states in Type IIA theory. In this section in 
particular, we shall consider the motion of an $M2$--brane as described by the Nambu--Goto 
action and we will show that the dynamics can be determined from the Born--Infeld action 
describing the $D2$--brane through a precise understanding of its Sen--Seiberg limit.

We start, in order to introduce notation\footnote{We will use the following index conventions. 
Indices in $M$--theory spacetime will be
$\mu,\nu,\dots$, and on the brane world--volume will be $\alpha,\beta,\dots$. Spatial indices
on the brane world--volume will be $a,b,\dots$, and spatial directions in IIA (or equivalently 
transverse directions in $M$--theory) will be $i,j,\dots$.},
by reviewing the standard light--cone description of 
$M$--theory in a flat background. Transverse coordinates will be denoted by
$X^i\, , \, i=1,\dots,9$, while 
light--cone ones by $X^{\pm} = 2^{-1/2} (X^0 \pm X^{10})$. Also, $\ell_P$ will denote the Planck 
length, and $R$ the light--like compactification radius in the $X^{-}$ direction. To describe the 
motion of an $M2$--brane (we shall always consider $M2$--branes which do not wrap the compact
direction), we take the world--volume of the 
membrane to be the product space $\mathbb{R} \times \Sigma$, where $\Sigma$ 
is a two dimensional surface with the topology of the brane. Coordinates on the world--volume are 
$\tau$ and $\sigma^{a}$ (with $a=1,2$) or, collectively, $\xi^{\alpha}$ (with $\alpha=0,1,2$). 
The motion of the surface is described by coordinate functions $X^{\mu}\,:\,\mathbb{R}\times \Sigma 
\rightarrow \mathbb{R}^{11}$ on the world--volume, and the correct dynamics is determined, at low
energy, by the minima of the membrane Nambu--Goto action,

$$
S = -T \int d^{3}\xi\, \sqrt{-\det \ g_{\alpha \beta} \ } \, ,
$$

\noindent
where $g_{\alpha \beta}=\partial_\alpha X^\mu \partial_\beta X_\mu$ is the induced metric, and the 
tension $T$ of the brane is given by

$$
T={1 \over (2\pi)^2 {\ell_P}^3} \, .
$$

\noindent
In order to simplify the highly non--polynomial Nambu-Goto action, we can use the
diffeomorphism invariance on the brane world--volume.
In fact, it is well known \cite{Goldstone-Hoppe,dhn} that the propagation of the $M2$--brane is most 
conveniently described in the light--front gauge, where the independent degrees of freedom are the 
transverse embedding coordinates $X^{i}$ and the remaining light-front coordinates $X^{\pm}$ are 
given by the gauge constraints $X^{+}=\tau$ and $\partial_{a}X^{-}=\dot{X}^{i}\partial_{a}X^{i}$. 
The dynamics of the system is then governed by the Lagrangian \cite{Goldstone-Hoppe,dhn}

\be\label{500}
L = \frac{T}{\sqrt{2}}\int \,d^{2}\sigma \,\,Q\left( \frac{1}{2}\dot{X}^{i}\dot{X}^{i}-
\frac{1}{2}\{X^{i},X^{j}\}^{2}\right) \, ,
\ee

\noindent
where $d^{2}\sigma \,Q$ is a fixed volume form on the manifold $\Sigma$ describing the brane and

$$
\{A,B\}=\frac{1}{Q}\varepsilon ^{ab}\partial _{a}A\partial _{b}B
$$

\noindent
is the Poisson bracket corresponding to the volume form. The overall normalization of the
action has been chosen so that, when considering static solutions of the equations of motion, the 
area element $Q\, d^2\sigma$ corresponds to the area element of the brane given by the embedding
in the transverse directions. More precisely, if

$$
h_{ab} = \partial_a X^i \partial_b X^i \, ,
$$

\noindent
then, for solutions such that $X^\pm = \tau$ and $\dot X^i = 0$, we have that
$Q=\sqrt{\det h_{ab}}$. More generally 

$$
Q = \sqrt\frac{\det h_{ab}}{\dot X^- - \frac{1}{2}\dot X^i \dot X^i} \, .
$$

\noindent
Finally, let us recall that, in order to solve the gauge constraint
$\partial_{a}X^{-}=\dot{X}^{i}\partial_{a}X^{i}$, we clearly have to restrict our attention to
solutions satisfying the constraint

\be\label{600}
\{ X^i , \dot{X}^i \} = 0 \, .
\ee

\noindent
We now wish to rederive the above description of the $M2$--brane, now starting from the known
duality between $M$--theory and Type IIA  \cite{Witten1}.
First, let us recall \cite{BFSS,Susskind,Seiberg,Sen} that $M$--theory 
compactified on the light--like circle $X^-\sim X^- + 2\pi R$ can be described as the
limit of $M$--theory compactified on a space--like circle of radius $\widetilde R$,
where we take the Sen--Seiberg limit $\widetilde R \rightarrow 0$.
Moreover, in order to match energy levels \cite{Seiberg,Sen},
we need to consider $M$--theory with a new Planck length
$\widetilde{\ell_P}$, where we keep the ratio $R/\ell_P^2 = \widetilde R/ \widetilde{\ell_P}^2$
constant. In the following, we shall refer to the space--like compactified theory as 
$\widetilde M$.
In order to describe the above limit in detail, we introduce a dimensionless parameter $\eta
\rightarrow 0$, and we then take

$$
\widetilde{\ell_{P}}=\eta ^{2}\ell_{P}\ \ , \ \ \ \ \ \  \widetilde{R}=\eta ^{4}R \, . 
$$

\noindent
If we denote by $\widetilde{X}^\mu$ the space--time coordinates in the $\widetilde M$--theory, we
have the relations

\beas
\widetilde{X}^0 &=& X^+ \, , \\
\widetilde{X}^{10} &\sim& \widetilde{X}^{10} + 2\pi\widetilde{R} \, , \\
\widetilde{X}^i &=& \eta ^{2} X^i \, .
\eeas

\noindent
Consider, in particular, a state in the theory $M$ with momentum $P^\mu$. The momentum $P^+$
corresponding to the compact direction $X^-$ will be quantized as

$$
P^+ = \frac{N}{R} \, .
$$

\noindent
Moreover, the corresponding state in the $\widetilde M$--theory will have momentum 
$\widetilde{P}^\mu$, where

\bea\label{400}
\widetilde{P}^0 &=& {N}/{\widetilde{R}}\, + P^+ \nonumber \\
\widetilde{P}^{10} &=& {N}/{\widetilde{R}} \nonumber \\
\widetilde{P}^i &=& \eta ^{-2} P^i \, .
\eea

Recall that $M$--theory compactified on a space--like circle is given by Type IIA string
theory at finite coupling \cite{Witten1}. Therefore, in order to describe the $\widetilde M$--
theory with $N$ units of momentum in the compact direction, one needs to consider Type IIA
with $N$ units of $D0$--brane charge and with the following string length and coupling

$$
\widetilde{\ell_s} = \eta \ell_s\ \ \ , \ \ \ \ \ \ \ \ \widetilde{g_s} = \eta^3 g_s \, ,
$$

\noindent
where we have introduced the constants

$$
{\ell_s}^2 = \frac{{\ell_P}^3}{{R}}\ \ \ , \ \ \ \ \ \ \ \ {g_s}
=\left( \frac{{\ell_P}}{{R}} \right)^{3/2} \, ,
$$

\noindent
related to the original light--like compactified $M$--theory. Finally, we introduce the constants

$$
\alpha ' = \ell_s^2\ \ \ , \ \ \ \ \ \ \ \epsilon = 2\sqrt{2}(2\pi \alpha ') \, ,
$$

\noindent
and the corresponding tilded quantities

$$
\widetilde{\alpha '} = \eta^2 \alpha '\ \ \ , \ \ \ \ \ \ \ \widetilde{\epsilon} = \eta^2 \epsilon \, .
$$

\noindent
In particular, the constant $\epsilon$ is introduced to make contact with \cite{CT,Cornalba}. It
plays the role of $\alpha '$, but it is more convenient to use, since it absorbs the factor of 
$2\pi$, which would otherwise appear explicitly in most of the equations that follow.

Let us now focus our attention on the description of brane states. In particular, let us
start by concentrating on static BPS branes in $M$--theory, given by embeddings
$X^\mu(\xi)$ which satisfy $X^\pm = \tau$ and $\dot{X}^{i}=0$, and which represent holomorphic
maps of the manifold $\Sigma$ in the transverse space $X^i$ (as described in detail in \cite{CT}).
Following the 
Sen--Seiberg limit described previously, we can consider the corresponding BPS brane states
within the theory $\widetilde M$, or better yet, within the corresponding Type IIA description.
In particular, let us first introduce the rescaled coordinates on the world-volume of the brane

\beas
&& \widetilde \xi^\alpha \sim (\wt\tau \, , \, \wt\sigma^a) \, , \\
&& \wt\tau = \tau\ \ \ , \ \ \ \ \ \ \ \ \wt\sigma^a = \eta^2 \sigma^a \, .
\eeas

\noindent
We are then considering, within the Type IIA description, a single static $D2$--brane 
given by the embedding
\beas
\widetilde{X}^0(\wt\xi) &=& X^+(\xi) = \tau \\
\widetilde{X}^i(\wt\xi) &=& \eta ^{2} X^i(\xi) \, .
\eeas

\noindent
Moreover, since in the original $M$--theory description the light--cone momentum $P^+$ of the 
brane is given by $P^+ = 2^{-1/2}\,P^0 = 2^{-1/2}\,TA\,$ ($A$ denotes the area of the membrane),
and since
$P^+ = N/R$, we must consider the sector of Type IIA with 

\be\label{100}
N= \frac{1}{\sqrt{2}}\, T A R
\ee

\noindent
units of $D0$--brane charge.

As is well known \cite{Leigh}, the motion of a $D2$--brane is described, in the low--energy limit,
by the Born--Infeld (BI) action

\begin{equation}\label{300}
\widetilde{S} =-\widetilde{T}\int d^{3}\widetilde{\xi }\sqrt{-\det \left( 
\widetilde{g}_{\alpha \beta }+2\pi \widetilde{\alpha }^{\prime }\widetilde{F}%
_{\alpha \beta }\right) } \, ,
\end{equation}

\noindent
where the tension of the brane is given by 

$$
\widetilde{T}={1 \over (2\pi)^2 \widetilde{g_s} \widetilde{\ell_s}^3}=\eta^{-6} T
$$

\noindent
and the induced metric by

$$
\wt g_{\alpha \beta}=-\wt\partial_\alpha\wt X^0 \wt\partial_\beta \wt X^0+
\wt\partial_\alpha\wt X^i \wt\partial_\beta \wt X^i \, .
$$

\noindent
This action, however, needs to be corrected 
at length scales comparable to the string length. These corrections take the form of terms 
with higher derivatives of the embedding functions and of the brane Maxwell field 
strength \cite{Ts}. Nonetheless, for the case we are now considering of static BPS states, 
one may only consider the BI action, since higher derivative terms do not change the nature of the
BPS solutions \cite{CM,Gibbons}.
In particular, we have already seen that the embedding functions are given by a holomorphic map
from $\Sigma$ to $X^i$. One then only needs to determine the correct Maxwell field on the membrane 
which describes the corresponding $M2$--brane state. Recalling that $\wt F_{\alpha\beta}$ measures 
$D0$--brane density on the $D2$--brane, and using equation (\ref{100}), we must have that

$$
\frac{1}{2\pi}\int_\Sigma \wt F_{12}\, d^2\wt\sigma =
\frac{\eta^4}{2\pi}\int_\Sigma \wt F_{12}\, d^2\sigma = 
N = \frac{1}{\sqrt{2}}\, TAR \, .
$$  

\noindent
It is not difficult to see that the correct solution to the BI action which satisfies the
above normalization is given by

$$
\wt F_{12} = \frac{1}{\eta^4\sqrt{2}}\frac{Q}{2\pi\alpha '} \, ,
$$

\noindent
where, as before, $Q\,d^2\sigma$ is the induced area element on the brane.
We then have that for the static BPS case, 

\begin{equation}\label{200}
2\pi \widetilde{\alpha }^{\prime }\widetilde{F}_{\alpha \beta }=\frac{Q}
{^{\eta ^{2}\sqrt{2}}}\left( 
\begin{array}{ccc}
0 & 0 & 0 \\ 
0 & 0 & 1 \\ 
0 & -1 & 0
\end{array}
\right) \, .
\end{equation}

Let us now consider fluctuations around the static solution, within the Type IIA description.
In order to make contact with the light--cone description of the $M2$--brane, we shall use 
part of the reparametrization invariance of the BI action. In particular, we shall keep
$\wt X^0 = \wt \tau = \tau$. Moreover, using time--dependent reparametrizations of the brane 
$\Sigma$ (which leave invariant the constraint $\wt X^0 =  \tau$), one can cancel any 
fluctuations  $\wt f_{\alpha\beta}$ of the $U(1)$ field--strength $\wt F_{\alpha\beta}+
\wt f_{\alpha\beta}$. This can be achieved by eliminating, first of all, the electric part
$\wt f_{0a}$ with a $\tau$--dependent reparametrization, using the fact that we are working in
a background with a large magnetic field (\ref{200}). One can then use a $\tau$--independent
reparametrization to eliminate the magnetic part $\wt f_{12}$. This can be done since, in 
$2$--dimensions, any two area elements (given in this case by $\wt F_{12}\,d^2\wt\sigma$ and by
 $(\wt F_{12}+ \wt f_{12}) \,d^2\wt\sigma$)
are always equivalent under reparametrization. Therefore, we can use diffeomorphism invariance
on the world--volume to bring any given configuration of the $D2$--brane into a configuration with
gauge field strength given by (\ref{200}) and with $\wt X^0 =  \tau$.

One then only needs to consider fluctuations in the embedding functions. We consider generic
transverse embedding functions

$$
\widetilde X^i(\wt\xi) = \eta ^{2} X^i(\xi) \, ,
$$

\noindent
recalling that we are always looking at fluctuations which
are finite from an $M2$--brane point of view, and are therefore finite in units of the 
Planck length. In units of the string length, these fluctuations are vanishingly small.
The induced metric $\wt g_{\alpha\beta}$ is given by

\begin{eqnarray*}
\widetilde{g}_{\alpha \beta } &=&\left( 
\begin{array}{cc}
-1+\widetilde{\partial }_{0}\widetilde{X}^{i}\widetilde{\partial }_{0}%
\widetilde{X}^{i} & \widetilde{\partial }_{0}\widetilde{X}^{i}\widetilde{%
\partial }_{b}\widetilde{X}^{i} \\ 
\widetilde{\partial }_{0}\widetilde{X}^{i}\widetilde{\partial }_{a}%
\widetilde{X}^{i} & \widetilde{\partial }_{a}\widetilde{X}^{i}\widetilde{%
\partial }_{b}\widetilde{X}^{i}
\end{array}
\right) = \\
&=&\left( 
\begin{array}{cc}
-1+\eta ^{4}\dot{X}^{i}\dot{X}^{i} & \eta ^{2}\dot{X}^{i}\partial _{b}X^{i}
\\ 
\eta ^{2}\dot{X}^{i}\partial _{a}X^{i} & \partial _{a}X^{i}\partial _{b}X^{i}
\end{array}
\right) =
\left( 
\begin{array}{cc}
-1+\eta ^{4}\dot{X}^{i}\dot{X}^{i} & \eta ^{2}\dot{X}^{i}\partial _{b}X^{i}
\\ 
\eta ^{2}\dot{X}^{i}\partial _{a}X^{i} & h_{ab}
\end{array}
\right) \, ,
\end{eqnarray*}

\noindent
where, as before,
$h_{ab} = \partial _{a}X^{i}\partial _{b}X^{i}$ is the spatial part of the induced metric.
We now wish to evaluate the BI action (\ref{300}) in the presence of fluctuations. On general
grounds one expects, in the limit of vanishingly small $\eta$, that

$$
\wt S = \int d{\tau }\left( -\frac{N}{\widetilde{R}}+L+o(\eta )\right) \, ,
$$

\noindent
corresponding to the fact that the energies in Type IIA are related to the light--cone 
momenta in $M$--theory by $\widetilde P^0 = {N}/{\widetilde{R}}\, + P^+$ (Eq. \ref{400}).
In the above equation, $L$ represents the Lagrangian in the $M$--theory limit, and
we shall show that it corresponds to the Lagrangian (\ref{500}). To compute the BI action,
we introduce the short--hand notation

$$
M_{\alpha\beta} = \widetilde{h}_{\alpha \beta }+2\pi \widetilde{\alpha }%
^{\prime }\widetilde{F}_{\alpha \beta } \, .
$$

\noindent
Given the choice of gauge previously described, we have that

$$
M_{\alpha \beta }=
\left( 
\begin{array}{ccc}
-1+\eta ^{4}\dot{X}^{i}\dot{X}^{i} & \eta ^{2}\dot{X}^{i}\partial _{1}X^{i}
& \eta ^{2}\dot{X}^{i}\partial _{2}X^{i} \\ 
\eta ^{2}\dot{X}^{i}\partial _{1}X^{i} & h_{11}
& \frac{Q}{^{\eta ^{2}\sqrt{2}}}+ h_{12} \\ 
\eta ^{2}\dot{X}^{i}\partial _{2}X^{i} & -\frac{Q}{^{\eta ^{2}\sqrt{2}}}%
+h_{21} & h_{22}
\end{array}
\right) \, ,
$$

\noindent
and, therefore, 

\begin{eqnarray*}
-\det M_{\alpha \beta } &=&(1-\eta ^{4}\dot{X}^{i}\dot{X}^{i})\left( \frac{%
Q^{2}}{2\eta ^{4}}+h\right) +o(\eta ) \\
&=&\frac{Q^{2}}{2\eta ^{4}}+h-\frac{Q^{2}}{2}\dot{X}^{i}\dot{X}^{i}+o(\eta ) \, ,
\end{eqnarray*}

\noindent
where $h = \det h_{ab}$. We can then evaluate the BI action in the $\eta\rightarrow 0$
limit (we drop, in the last lines, terms of order $\eta$),

\begin{eqnarray*}
\widetilde{S} &=& -\widetilde{T}\int d^{3}\widetilde{\xi }\sqrt{-\det 
M_{\alpha \beta }}\\
&=&-\frac{1}{\eta ^{2}}T\int d^{3}\xi \sqrt{\frac{Q^{2}}{2\eta
^{4}}+h-\frac{Q^{2}}{2}\dot{X}^{i}\dot{X}^{i}+o(\eta )} \\
&=&-\frac{1}{\eta ^{4}\sqrt{2}}T\int d^{3}\xi \,\ Q\left( 1+\frac{\eta ^{4}}{%
Q^{2}}h-\frac{1}{2}\eta ^{4}\dot{X}^{i}\dot{X}^{i}+o(\eta ^{5})\right)  \\
&=& \int d\tau \left( -\frac{TA}{\eta ^{4}\sqrt{2}} +\frac{T}{\sqrt{2}}\int
d^{2}\sigma \,\ \left( \frac{1}{2}Q\dot{X}^{i}\dot{X}^{i}-\frac{h}{Q}\right) \right) \\
&=&  \int d\widetilde{\tau }\left( -\frac{N}{\widetilde{R}}+L\right) \, ,
\end{eqnarray*}

\noindent
where

$$
L = \frac{T}{\sqrt{2}}\int
d^{2}\sigma \,\ \left( \frac{1}{2}Q\dot{X}^{i}\dot{X}^{i}-\frac{h}{Q}\right) \, .
$$

\noindent
Using the fact that

$$
h = \frac{1}{2} Q^2 \{X^i,X^j\}^2 \, ,
$$

\noindent
we recover the Lagrangian (\ref{500}).

To conclude this section we would like to derive, within the Type IIA description, the 
constraint (\ref{600}) which is an integral part of the light--cone Lagrangian (\ref{500}).
Recall that we have used diffeomorphism invariance of the BI action to gauge away the 
fluctuations in the $U(1)$ field--strength. Invariance of the action $\wt S$ under
infinitesimal fluctuations of $\wt F_{\alpha\beta}$ should then be reinterpreted as a 
constraint on the allowed configurations of the Lagrangian system under consideration.
In particular, we shall show that invariance of $\wt S$ under infinitesimal fluctuations of
the electric field will yield, in the $\eta\rightarrow 0$ limit, the constraint (\ref{600})
on the embedding functions. 
Start by considering a generic fluctuation $\wt f_{\alpha\beta}=\partial_\alpha a_\beta -
\partial_\beta a_\alpha$. Then 

\beas
\widetilde{S}+\delta \widetilde{S} &\propto& \int d^{3}\xi 
\sqrt{-\det \left( M_{\alpha \beta }+\wt f_{\alpha \beta }\right) } \\
\delta \widetilde{S} &\propto& \int d^{3}\xi \sqrt{-\det
M_{\alpha \beta }}\,\ M^{\alpha \beta }\wt f_{\beta \alpha } \, .
\eeas

\noindent
It is easy to see that, to leading order in $\eta$, the inverse matrix $M^{\alpha\beta}$ is
given by (we only show the elements of the inverse matrix which will enter in the computation
that follows)

\begin{equation}\label{700}
M^{\alpha \beta }=\frac{1}{\det M_{\alpha \beta }}\left( 
\begin{array}{ccc}
\cdots  & -\frac{Q}{\sqrt{2}}\dot{X}^{i}\partial _{2}X^{i} & \frac{Q}{\sqrt{2%
}}\dot{X}^{i}\partial _{1}X^{i} \\ 
\frac{Q}{\sqrt{2}}\dot{X}^{i}\partial _{2}X^{i} & \cdots  & \cdots  \\ 
-\frac{Q}{\sqrt{2}}\dot{X}^{i}\partial _{1}X^{i} & \cdots  & \cdots 
\end{array}
\right) +\cdots \, .
\end{equation}

\noindent
Let us now concentrate on the precise form of the infinitesimal fluctuation $\wt f_{\alpha\beta}$.
First of all we can use $U(1)$ gauge invariance to move to the $a_0 = 0$ gauge. Moreover, since
we are considering purely electric fluctuations, we have that $\partial_1 a_2 =
\partial_2 a_1$. Therefore $a_a = \partial_a f$. If we let $\lambda = \partial_0 f$,
we have that

$$
\wt f_{0a} = \partial_a \lambda \, .
$$

\noindent
We can then use equation (\ref{700}) to show that

\begin{eqnarray*}
\delta \widetilde{S} &\propto& \int d^{3}\xi \frac{1}{\sqrt{-\det M_{\alpha \beta }%
}}\,\frac{Q}{\sqrt{2}}\ \left( \partial _{1}\lambda \,\dot{X}^{i}\partial
_{2}X^{i}-\partial _{2}\lambda \,\dot{X}^{i}\partial _{1}X^{i}\right)  \\
&\propto& \int d^{3}\xi \left( \partial _{1}\lambda \,\dot{X}^{i}\partial
_{2}X^{i}-\partial _{2}\lambda \,\dot{X}^{i}\partial _{1}X^{i}\right)  \\
&\propto&\int d^{3}\xi \left( \partial _{1}X^{i}\partial _{2}\dot{X}%
^{i}-\,\partial _{2}X^{i}\,\partial _{1}\dot{X}^{i}\right) \lambda \, .
\end{eqnarray*}

\noindent
Since $\lambda$ is arbitrary, we recover the constraint,

$$
\partial _{1}X^{i}\partial _{2}\dot{X}^{i}-\,\partial _{2}X^{i}\,\partial
_{1}\dot{X}^{i}=0=\{X^{i},\dot{X}^{i}\} \, ,
$$

\noindent
as given in equation (\ref{600}).

\sect{Higher Derivative Corrections to the Born--Infeld and the Star Product}

This final section is devoted to a qualitative discussion on the higher--order corrections
to the Born--Infeld (BI) action and to the analysis of their Sen--Seiberg limit. The limit will be 
reinterpreted as a Matrix theory action living on the world--volume of the $M2$-brane.

Let us start by recalling the Matrix theory Lagrangian \cite{BFSS,Susskind}

\be\label{1000}
L=\frac{1}{R} {\rm Tr}\left( \frac{1}{2}\dot{X}^{i}\dot{X}^{i}+{\frac{2}{\epsilon
^{2}}}[X^{i},X^{j}]^{2}\right) \, .
\ee

\noindent
It is given by the dimensional reduction of $10$--dimensional $U(N)$ SYM--theory to
$(0+1)$--dimensions. It is the low--energy limit of the BI action describing the degrees
of freedom of $N$ $D0$--branes and it has been conjectured \cite{BFSS,Susskind} to be an exact
non--perturbative description of light--cone compactified $M$--theory, in the sector
with $N$ units of momentum in the compact direction. The Lagrangian (\ref{1000}) should
be supplemented with the constraint

$$
[X^i, \dot X^i ] = 0 \, ,
$$

\noindent
which comes from the choice of temporal gauge in the SYM theory. The conjecture itself
can again be understood in terms of the Sen--Seiberg limit described in the previous
section \cite{Seiberg,Sen}.

The Lagrangian (\ref{1000}) should contain all of the physics of $M$--theory. In 
particular, it should describe sectors of the theory with finite $M2$--brane charge.
Let us consider, to start, a specific BPS $M2$--brane, described at the classical level
by a holomorphic curve in the transverse space. To be definite,
introduce the complex coordinates $s=
\sigma^1+i\sigma^2$ and $Z_1=X^1+iX^2$, $Z_2=X^3+iX^4$, $\dots$, so that the coordinate functions
$Z_A$ are analytic in $s$. It was shown in \cite{CT} that we can associate 
to each such curve a specific BPS state in Matrix theory, which gives a
matrix representation of the given holomorphic brane. We shall not review here 
the construction given in \cite{CT}, but we will quickly describe what is needed
for the discussion of this section. In \cite{CT}, given a fixed BPS brane, a map
${\cal Q}$ was constructed which associates to each function $A$ on the brane $\Sigma$
a matrix ${\cal Q}(A)$ (or better an operator on a given Hilbert space). The map $\cal Q$
has the property that, given any two functions $A,B$ on $\Sigma$, on has

\be\label{1100}
{\cal Q}(A) {\cal Q}(B) = {\cal Q} (A \star B) + o(\epsilon^\infty) \, ,
\ee

\noindent
where $\star$ is a specific associative star product associated with the given curve,
which we shall describe in detail in what follows. As indicated above, the equality in equation 
(\ref{1100}) is asymptotic in $\epsilon$, and is valid only up to corrections which vanish 
faster then any power of $\epsilon$. In this paper we shall not be interested in non--perturbative
effects in $\epsilon$ and will therefore ignore such corrections, treating the equality
in (\ref{1100}) as exact. In this case, we can speak interchangeably of functions on the brane together
with a given star product, or of matrices together with matrix multiplication. In order to make
contact with the membrane theory of the last section, it is clear that the description in terms
of functions on the brane is the preferable one, and we will, from now on, consider matrices as
functions on $\Sigma$ (with the product $\star$). 

Let us now describe star products in detail. A product $A\star B$ is given as a formal power
series $\sum_{n=0}^\infty \epsilon^n S_n(A,B)$ in $\epsilon$, where at each order the coefficient
$S_n(A,B)$ is a local bilinear in $A$ and $B$, built out of the derivatives of $A$ and $B$ up to finite 
order. In particular, we will have,
 
\bea\label{2000}
S_0(A,B) &=& AB \\
S_1(A,B) - S_1(B,A) &=& \frac{1}{2i} \{ A,B \} \, . \nonumber
\eea

\noindent
As is well known \cite{bayen,lecomte,fedosov,maxim}, the condition of associativity puts stringent constraints on the 
explicit form of the coefficients $S_n$.

We shall further constrain the star product by demanding that it preserves the BPS nature of
the holomorphic brane which we are describing. In particular, the supersymmetry 
transformations of the Lagrangian (\ref{1000}) are known and simple, and one can show 
\cite{CT} that the correct BPS conditions which are appropriate for a supersymmetric
membrane state are given by

\bea\label{1200}
&& [Z_A, Z_B] = [Z_A^\dagger, Z_B^\dagger] = 0 \nonumber \\
&& \sum_A \, [Z_A, Z_A^\dagger] = -\epsilon \, .
\eea

\noindent
Imposing the above equations gives an extra constraint on the star product on the brane.
It is carefully shown in \cite{CT} that the conditions above are solved by the following 
product

$$
A \star B = \sum_{n=0}^{\infty} \,\epsilon^n\, Q_1 \cdots Q_n \, \left( {1\over Q_n} \partial \cdots 
{1\over Q_1} \partial A\right)\,
\left( {1\over Q_n} \bar{\partial} \cdots {1\over Q_1} \bar{\partial} B\right) \, .
$$

\noindent
The coefficients $Q_n$ are $(1,1)$ tensors on the brane, which can be computed recursively
starting from $Q_1$ with the formula,

$$
Q_n = Q_{1} + Q_{n-1} + \epsilon\, \partial\bar\partial \log (Q_1\cdots Q_{n-1}) \, .
$$

\noindent
The tensor $Q_1$ can, in turn, be computed perturbatively in $\epsilon$ by imposing equation
(\ref{1200}) and is given by

$$
Q_1 = -Q + \frac{\epsilon}{2}\partial\bar\partial \log Q + o(\epsilon^2) \, .
$$

Let us also briefly discuss traces of operators. The major fact is that, again asymptotically
in $\epsilon$, they can be computed as integrals over the surface $\Sigma$,

\be\label{2100}
{\rm Tr} ({\cal Q}(A)) = \frac{1}{\epsilon}\int_\Sigma\ \sum_{n=0}^\infty \epsilon^n \mu_n\, A \, ,
\ee

\noindent
where

$$
\mu_0 = \frac{1}{\pi\epsilon} Q\, d^2\sigma \, .
$$

We now arrive to the fundamental point of this paper. One starts with a static BPS solution
and considers the physics of the fluctuations, as governed by the Lagrangian (\ref{1000}).
If one considers the fluctuations not as matrices but, in the spirit of what we have just
discussed, as functions on the brane $\Sigma$, and if morover one replaces matrix products by star 
products and traces with integrals, one can rewrite the action (\ref{1000}) as a field theory
on $\Sigma \times {\bf R}$ governing the fluctuations of the brane itself. Using equations
(\ref{2000},\ref{2100}), we have

\begin{eqnarray*}
A\cdot B &\rightarrow &AB + \cdots\\
\lbrack ,] &\rightarrow &\frac{\varepsilon }{2i}\{,\} + \cdots \\
\rm{Tr} &\rightarrow &\frac{1}{\pi \varepsilon }\int d^{2}\sigma \,Q +\cdots \, ,
\end{eqnarray*}

\noindent
Therefore the Matrix theory action (\ref{1000}) has an $\epsilon$ expansion which starts
as

$$
L = \frac{1}{\pi\epsilon R}\int \,d^{2}\sigma \,\,Q\left( \frac{1}{2}\dot{X}^{i}\dot{X}^{i}-
\frac{1}{2}\{X^{i},X^{j}\}^{2}\right) \, + \cdots \, .
$$

\noindent
The above reproduces the action (\ref{500}), if one recalls that $\pi\epsilon R T = \sqrt{2}$.
Moreover, the full gauge constraint $[\dot X^i , X^i]=0$ for the Matrix theory Lagrangian does 
imply, to leading order in $\epsilon$, the constraint $\{ \dot X^i, X^i \} =0$ discussed in the last 
section.

Let us analyze higher order terms. From the point of view of the $D2$--brane action, one must 
understand the structure of the terms which involve higher derivatives of the embedding 
coordinates (curvature terms) and of the gauge field strength. These terms must be analyzed
in the limit of very large background constant field--strength, as is clear from 
equation (\ref{200}), even though we can work order by order in derivatives. Since we are working
in the limit of vanishing string coupling, one can analyze open--string disk diagrams, where
the underlying CFT is that of a string ending on a brane with a constant field--strength. 
Derivative corrections to the effective action can then be analyzed by computing standard string 
amplitudes, in the presence of the given background. At first sight we expect the derivative
corrections to the BI action to diverge in the Sen--Seiberg limit. Indeed, we
expect for example terms containing the curvature of the brane, measured in units of the
string length. Since the characteristic length scale of the brane is proportional, 
in the Sen--Seiberg limit, to the Planck length, and since the string length $\wt\ell_s\sim\eta$  
is large compared to $\wt\ell_P\sim\eta^2$, one expects curvature terms to dominate in the limit. 
However, in this argument we are neglecting the fact that we are working in a background of
large gauge field strength $2\pi\wt\alpha ' \wt F_{\alpha\beta} \sim 1/\eta^2$. The curvature 
corrections can then be altered by inverse powers of the constant background gauge field strength
(as can be seen by looking at the propagator in the boundary CFT in the presence of a background
$F$), thus yielding a finite result. 

It is therefore natural to conjecture the following.
{\it The Sen--Seiberg limit of the action describing the $D2$--brane in Type IIA string theory can 
be resummed to all orders in $\alpha '$. The full sum can be then re--written as the action 
(\ref{1000}) describing Matrix theory, where one interprets matrix multiplication as an 
associative star product on the brane world--volume. The star product itself must,
for consistency reasons, satisfy the conditions (\ref{2000}) and (\ref{1200}).}

Let us conclude by commenting on some questions of uniqueness. In section 2 we carefully constructed
the Sen--Seiberg limit by imposing the constraint (\ref{200}). This clearly depends on the exact 
definition of the $U(1)$ gauge field living on the brane. In particular, recall that one can 
always do a field redefinition of the gauge field by adding gauge--invariant quantities. Therefore, the
precise form of the Sen--Seiberg limit depends on the chosen definition. It is then natural to 
expect that different definitions will yield in the limit actions written in terms of 
different star products. All these actions should be equivalent, since they can be resummed
to yield the same Matrix theory action. Moreover, any consistent star product will have 
to be associative and will have to satisfy conditions (\ref{2000}) and (\ref{1200}).

\vs{5}
\noindent
{\bf Acknowledgments}
We would like to thank W.~Taylor for helpful discussions and comments, and also M.~Douglas, 
A.~A.~Tseytlin and N.~Seiberg for useful comments. One of us (RS) was supported in part by funds 
provided by the U.S. Department of Energy (D.O.E.) under cooperative research agreement 
$\sharp$DE-FC02-94ER40818, in part by Funda\c c\~ao Calouste Gulbenkian (Portugal), and in part by 
Funda\c c\~ao Luso--Americana para o Desenvolvimento (Portugal).

\eject

\newcommand{\NP}[1]{Nucl.\ Phys.\ {\bf #1}}
\newcommand{\PL}[1]{Phys.\ Lett.\ {\bf #1}}
\newcommand{\CMP}[1]{Comm.\ Math.\ Phys.\ {\bf #1}}
\newcommand{\PR}[1]{Phys.\ Rev.\ {\bf #1}}
\newcommand{\PRL}[1]{Phys.\ Rev.\ Lett.\ {\bf #1}}
\newcommand{\PTP}[1]{Prog.\ Theor.\ Phys.\ {\bf #1}}
\newcommand{\PTPS}[1]{Prog.\ Theor.\ Phys.\ Suppl.\ {\bf #1}}
\newcommand{\MPL}[1]{Mod.\ Phys.\ Lett.\ {\bf #1}}
\newcommand{\IJMP}[1]{Int.\ Jour.\ Mod.\ Phys.\ {\bf #1}}
\newcommand{\JP}[1]{Jour.\ Phys.\ {\bf #1}}
\newcommand{\JMP}[1]{Jour.\ Math.\ Phys.\ {\bf #1}}
\newcommand{\AP}[1]{Annals\ of\ Phys.\ {\bf #1}}
\newcommand{\LMP}[1]{Lett.\ Math.\ Phys.\ {\bf #1}}
\newcommand{\JDG}[1]{Jour.\ Diff.\ Geo.\ {\bf #1}}
\newcommand{\ATMP}[1]{Adv.\ Theor.\ Math.\ Phys.\ {\bf #1}}
\newcommand{\JHEP}[1]{JHEP\ {\bf #1}}

\end{document}